\documentclass[aps,pre,preprint,groupedaddress,showpacs,amsmath]{revtex4}
\usepackage{graphicx}
\usepackage{bm}
\begin{document}  

\title{Helical filaments with varying cross section radius}
  
\author{Alexandre Fontes da Fonseca \footnote{Present address:
Department of Chemistry, Rutgers, The State University of New Jersey,
Piscataway, New Jersey, 08854-8087 USA}}

\email[]{afonseca@if.usp.br}
\author{Coraci P. Malta}
\email[]{coraci@if.usp.br}


\affiliation{
Instituto de F\'{\i}sica, Universidade de S\~ao Paulo,
USP\\ Caixa Postal 66318, 05315-970, S\~ao Paulo, Brazil}

\date{\today}

\begin{abstract}

The tridimensional configuration and the twist density of helical rods
with varying cross section radius are studied within the framework of
the Kirchhoff rod model. It is shown that the twist density increases
when the cross section radius decreases. Some tridimensional
configurations of helix-like rods are displayed showing the effects of
the nonhomogeneity considered here. Since the helix-like solutions of
the nonhomogeneous rods do not present constant {\it curvature} and
{\it torsion} a set of differential equations for these quantities is
presented. We discuss the results and possible consequences.

\end{abstract}

  
\maketitle  


Helical filaments are tridimensional structures universally found in
Nature. They can be seen in very small sized systems, as
biomolecu\-les~\cite{tamar} and bacterial fibers~\cite{wolge}, and in
macroscopic ones, as ropes, strings, and climbing
plants~\cite{alain,foot1,tyler}. All these objects have in common the
fact that the mathematical geometric properties of the 3D-space curve
related to their axis, namely the {\it curvature}, $k_F$, and the {\it
torsion}, $\tau_F$, are constant~\cite{nize}.

The so-called Kirchhoff rod model has been proved to be a good
framework to study the statics~\cite{nize,vander} and
dynamics~\cite{goriely} of long, thin and inextensible elastic
rods~\cite{kirch,dill}. The applications of the Kirchhoff model range
from Biology~\cite{tamar,col1,tyler} to Engineering~\cite{sun}. In
these cases, the rod or filament is considered as being homogeneous,
but the case of nonhomogeneous rods have been also considered in the
literature. It has been shown that nonhomogeneous Kirchhoff rods may
present spatial chaos \cite{holmes,davies} and that helical
transitions occur in the tridimensional configurations of rods with
periodic variation of the Young's modulus~\cite{fonseca0}. A
comparison between homogeneous and nonhomogeneous rods subject to
given boundary conditions and mechanical parameters was performed by
da Fonseca and de Aguiar in~\cite{fonseca1}. The effects of a
nonhomogeneous mass distribution in the dynamics of unstable closed
rods have been analyzed by Fonseca and de
Aguiar~\cite{fonseca2}. Goriely and McMillen~\cite{alain2} studied the
dynamics of cracking whips~\cite{whip} and Kashimoto and
Shiraishi~\cite{moto} studied twisting waves in inhomogeneous rods.

Here, using the Kirchhoff model, we shall present the results for the
equilibrium solutions of nonhomogeneous rods with varying cross
section radius and no intrinsic curvature. Only the solutions
classified by Nizette and Goriely~\cite{nize} as being helical will be
considered: the straight rod, the twisted planar ring and the
helix. We shall show that the twist density varies along the rod
inversely proportional to the fourth power of the radius of the cross
section. Also, it will be seen that the curvature, $k_F$, and the
torsion, $\tau_F$, are not constant for the helix-like solutions with
nonhomogeneous cross section radius.

The motivations for this work are: i) the study of failure or rupture
of cables~\cite{azevedo,sayed}. For example, it was shown that
shoreline anchor rods rupture at the region where the rod diameter
diminishes due to corrosion~\cite{sayed}. The fact that, for twisted
rods, the twist density increases in the regions where the rod
diameter decreases can be related to the onset of the failure; ii) the
shape of some climbing plants have filamentary helical structures
(spring-like tendrils) whose radius and pitch are not
constant~\cite{alain}. Such a tridimensional configuration, with the
radius and the pitch varying along the rod, will be shown to be a
possible solution of the Kirchhoff model for the nonhomogeneous
rod. Other motivations are related to defects~\cite{kronert},
distortions~\cite{geetha} and the rule of twisting~\cite{turner} in
biological molecules.

The static Kirchhoff equations, in scaled variables, for rods with
circular cross section and no intrinsic curvature are given by:
\begin{equation}
\label{kir1}
\begin{array}{ll}
\mathbf{F}'=0 \; , \\
\mathbf{M}'+\mathbf{d}_{3}\times\mathbf{F}=0 \; , \\
\mathbf{M}=I(s)\,k_1\,\mathbf{d}_1+I(s)\,k_2\,\mathbf{d}_2+\Gamma\,
I(s)\,k_3\,\mathbf{d}_3 \; , 
\end{array}
\end{equation}
where $s$ is the arc-length of the rod, the prime $'$ denotes
differentiation with respect to $s$ and the vectors $\mathbf{F}$ and
$\mathbf{M}$ are the resultant force, and corresponding moment with
respect to the axis of the rod, respectively, at each cross
section. $\mathbf{d}_i$, $i=1,2,3$, compose the director basis with
$\mathbf{d}_3$ chosen to be the vector tangent to the axis of the rod
and $\mathbf{d}_1$ and $\mathbf{d}_2$ lie in the plane of the cross
section. $k_i$ are the components of the twist vector, $\mathbf{k}$,
that controls the variations of the director basis along the rod
through the relation $\mathbf{d}'_i=\mathbf{k}\times
\mathbf{d}_i$. $k_1$ and $k_2$ are related to the curvature
($k_F=\sqrt{k^{2}_{1}+k^{2}_{2}}$) and $k_3$ is the twist density of
the rod. $\Gamma =2\mu / E$ is the adimensional elastic parameter,
with $\mu$ and $E$ being the shear and the Young's moduli,
respectively. $I(s)$ is the variable moment of inertia that is related
to the radius of the cross section through the relation
$I(s)=R^{4}(s)$ (valid in scaled units). Writing the resultant force
$\mathbf{F}$ in the director basis,
$\mathbf{F}=f_1\mathbf{d}_1+f_2\mathbf{d}_2+f_3\mathbf{d}_3$, the
equations (\ref{kir1}) give six differential equations for the
components of the resultant force and twist vector:
\begin{subequations}
\label{es1}
\begin{eqnarray}
f'_1-f_2\, k_3+f_3\, k_2=0 \; , \label{a}  \\
f'_2+f_1\, k_3-f_3\, k_1=0 \; , \label{b}  \\
f'_3-f_1\, k_2+f_2\, k_1=0 \; , \label{c}  \\
(I(s)\, k_1)'+(\Gamma-1)\, I(s)\, k_2\, k_3-f_2=0 \; , \label{d} \\
(I(s)\, k_2)'-(\Gamma-1)\, I(s)\, k_1\, k_3+f_1=0 \; , \label{e} \\
(\Gamma\ I(s)\ k_3)'=0 \; . \label{f} 
\end{eqnarray}
\end{subequations}
Since $I(s)=R^{4}(s)$, the equation (\ref{f}) shows that the twist
density $k_3$ is inversely proportional to $R^{4}(s)$. Therefore,
$k_3$ is not constant for nonhomogeneous cases. On the other hand, the
component $M_3=\Gamma\,I\,k_3$ of the moment in the director basis
(also called {\it torsional moment}), is a constant along the rod.

In order to look for helical solutions of the eqs. (\ref{es1}) the
components of the twist vector $\mathbf{k}$ are expressed as follows:
\begin{equation}
\label{hel}
k_1=k_F\sin\xi \; , \;
k_2=k_F\cos\xi \; , \;
k_3=\xi'+\tau_F \; , 
\end{equation}
where $k_F$ and $\tau_F$ are the curvature and torsion of the rod,
respectively. In the homogeneous case $k_F$ and $\tau_F$ are constant,
and $\xi=(k_3-\tau_F)s$~\cite{nize}.

We shall consider the following cases: the straight rod, the twisted
planar ring and the general helix-like rod.

i) The straight rod: $k_F=\tau_F=0$. 

\begin{figure}[ht]
  \begin{center}
  \includegraphics[height=35mm,width=95mm,clip]{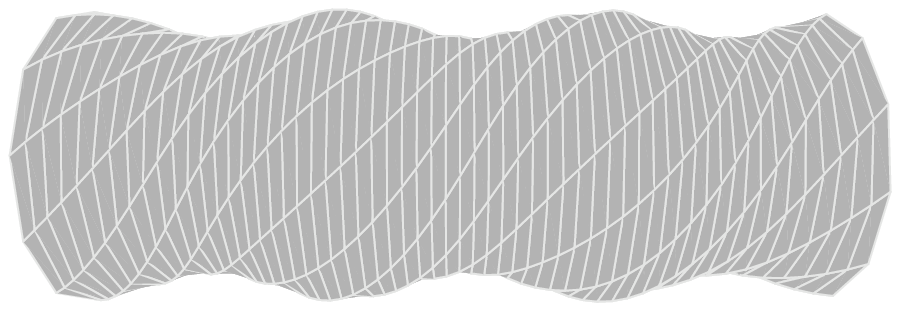}
  \includegraphics[height=55mm,width=100mm,clip]{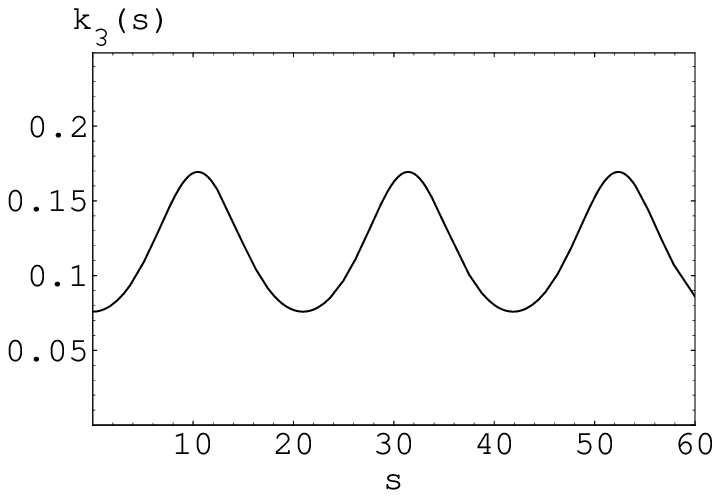}
  \end{center} 
  \caption{Top: a twisted straight rod with varying
  radius of the cross section. Bottom: the twist density $k_3$ as
  function of the arc-length $s$.}
\label{fig1}
\end{figure}

From eq. (\ref{hel}), $k_1=k_2=0$ and from eqs. (\ref{es1}),
\begin{equation}
\begin{array}{ll}
\label{k3}
k_3(s)=\frac{\mbox{Constant}}{I(s)}=\frac{M_3}{\Gamma\, R^{4}(s)} \; ,\\ 
f_1=f_2=0 \; \mbox{and} \; f_3\equiv T=\mbox{Constant} \; .
\end{array}
\end{equation}
$T$ is the tension applied to the rod. Figure \ref{fig1} shows the
twist density $k_3(s)$ for a straight rod with the cross section
radius varying as
\begin{equation}
\label{Rs}
R(s)=1+0.1\cos(0.3s) \; ,
\end{equation}
in scaled units. The mechanical parameters used to obtain the rod
displayed in Figure \ref{fig1} were $M_3=10^{-1}$ (scaled units) and
$\Gamma =0.9$.

ii) The twisted planar ring: $k_F=$ Constant and $\tau_F=0$.

The components of the twist vector for the twisted planar ring are:
\begin{equation}
\label{tpr1}  
\begin{array}{ll}
k_1=k_F\sin\xi \; , \;
k_2=k_F\cos\xi \; , \;
\xi '=k_3=\frac{M_3}{\Gamma\, R^{4}(s)} \; .
\end{array}
\end{equation}
Substituting the eqs. (\ref{tpr1}) in eqs. (\ref{es1}) shows that the
twisted planar ring is a possible equilibrium solution only if the
cross section radius is of the form:
\begin{equation}
\label{soltpr}
R(s)=(A_0\cos(k_F\,s)+B_0\sin(k_F\,s)+
C_I/k_F^{2})^{\frac{1}{4}} \; ,
\end{equation}
where $A_0$, $B_0$ and $C_I$ are constants. 

{\bf Remark}: considering $\tau_F=0$ and assuming that $k_F$ is a
function of $s$ (instead of being a constant) there exist no solutions
for eqs. (\ref{es1}). So, the existence of planar solution requires
$k_F=$ Constant.

iii) The general helix-like rod: $k_F=k_F(s)$ and $\tau_F=\tau_F(s)$.

In this case, the eqs. (\ref{hel}) become:
\begin{equation}
\label{hel_s}
k_1=k_F(s)\sin\xi \; , \;
k_2=k_F(s)\cos\xi \; , \;
k_3=\xi'+\tau_F(s) \; . 
\end{equation}
Substituting eq. (\ref{hel_s}) in eqs. (\ref{es1}), extracting $f_1$
and $f_2$ from eqs. (\ref{e}) and (\ref{d}), respectively,
differentiating them with respect to $s$ and substituting in
eqs. (\ref{a}) and (\ref{b}) gives a set of differential equations for
$k_F(s)$, $\tau_F(s)$ and $f_3(s)$:
\begin{equation}
\label{dif}
\begin{array}{ll}
[(I(s)\,k_F(s))(\Gamma\,k_3(s)-\tau_F(s))]'-(I(s)\,k_F(s))'\,\tau_F(s)
=0 \; ,  \\
(I(s)\,k_F(s))''+I(s)\,k_F(s)\,\tau_F(s)(\Gamma\,k_3(s)-\tau_F(s))
-f_3(s)\,k_F(s)=0 \; ,  \\
(I(s)\,k_F(s))'\,k_F(s)+f'_3(s)=0 \; .   
\end{array}
\end{equation}
These differential equations are nonlinear and depend on $R(s)$
through $I(s)$.

\begin{figure}[ht]
  \begin{center}
  \includegraphics[height=35mm,width=30mm,clip]{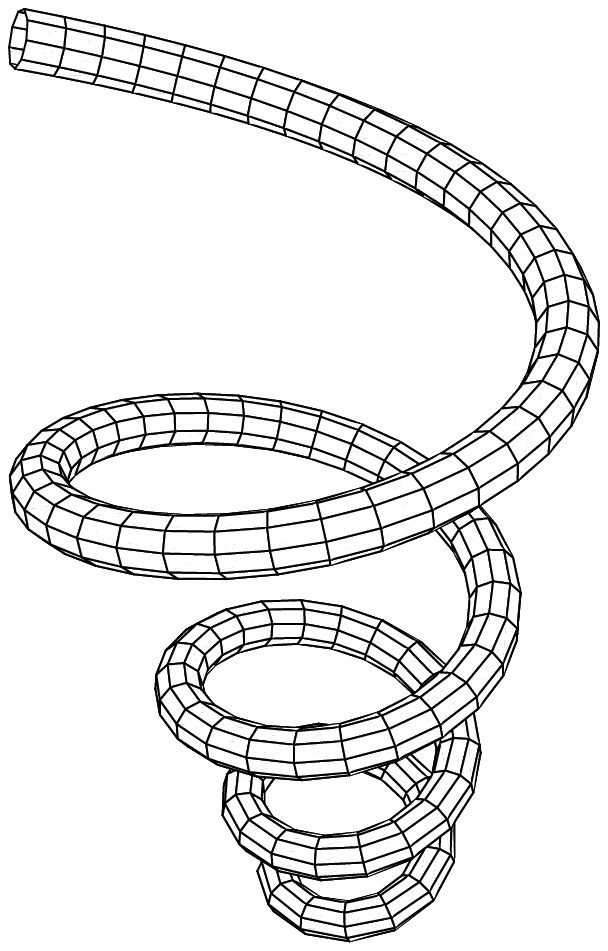}
  \includegraphics[height=35mm,width=55mm,clip]{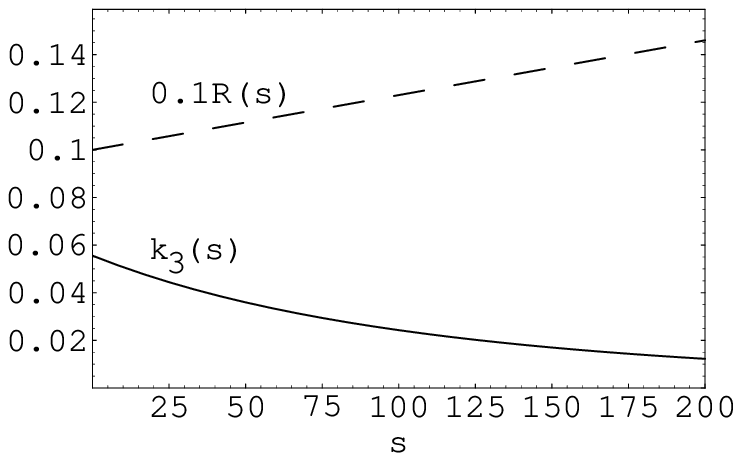}
  \end{center} 
  \caption{Left: helix-like rod having the radius of the cross section
  varying linearly with $s$ (eq. (\ref{RsHel1})). Right: $0.1\,R(s)$
  (dashed line) and $k_3(s)$ (full line).}
\label{fig2}
\end{figure}

Figure \ref{fig2} shows a helix-like rod for the following linear
variation of the radius of the cross section:
\begin{equation}
\label{RsHel1}
R(s)=1+0.0023\,s \; .
\end{equation}
The mechanical parameters, in scaled units, are $M_3=0.05$, $\Gamma
=0.9$, $k_F(0)=0.05$, $\tau_F(0)=0.24$ and $f_3(0)=0$. 
In the figure \ref{fig2} we display $k_3(s)$ (full line) and
$0.1\,R(s)$ (dashed line). We can see that the radius and pitch of the
helix-like tridimensional configuration displayed on the left of
figure \ref{fig2} are not constant. 

\begin{figure}[ht]
  \begin{center}
  \includegraphics[height=40mm,width=75mm,clip]{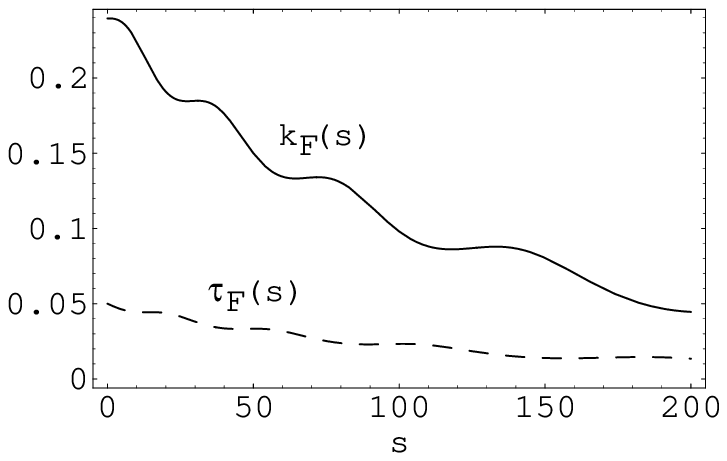} 
  \end{center}
  \caption{Geometric parameters: curvature, $k_F(s)$ (full line), and
  torsion, $\tau_F(s)$ (dashed line), for the equilibrium solution
  shown in the figure \ref{fig2}.}
\label{fig3}
\end{figure}

Figure \ref{fig3} shows the numerical solution for the curvature
$k_F(s)$ (full line) and torsion $\tau_F(s)$ (dashed line) for the
helix-like rod shown in the figure \ref{fig2}. Despite the simplicity
of its tridimensional shape (figure \ref{fig2}, on the left) $k_F(s)$
and $\tau_F(s)$ are not simple functions of the arclength $s$, showing
the nonlinear characteristic of the system.

\begin{figure}[ht]
  \begin{center}
  \includegraphics[height=35mm,width=30mm,clip]{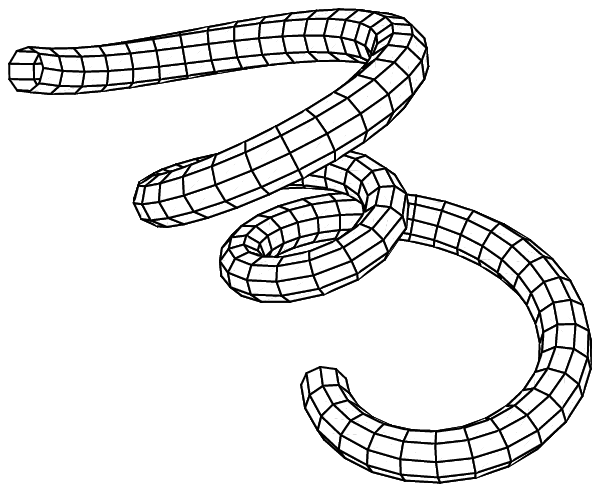}
  \includegraphics[height=35mm,width=55mm,clip]{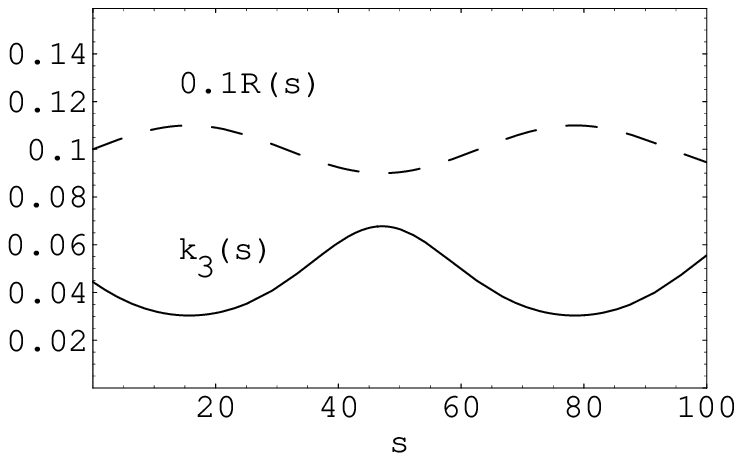}
  \end{center} 
  \caption{Left: helix-like rod having the radius of the cross section
  varying periodically with $s$ (eq. (\ref{RsHel2})). Right: $0.1\,R(s)$
  (dashed line) and $k_3(s)$ (full line).}
\label{fig4}
\end{figure}

\begin{figure}[ht]
  \begin{center}
  \includegraphics[height=40mm,width=75mm,clip]{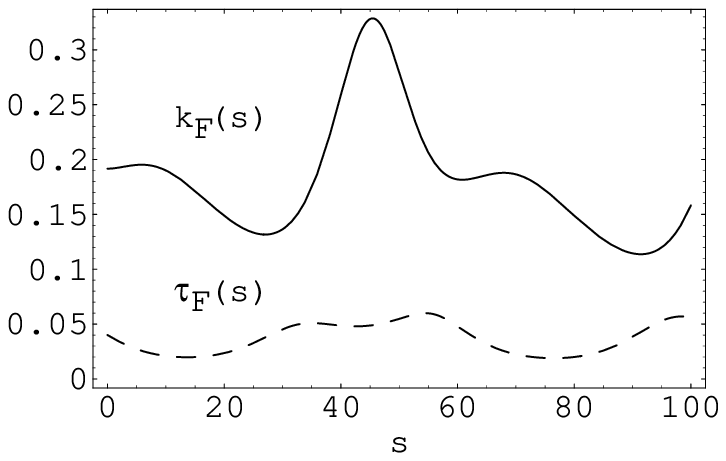}
  \end{center} \caption{Geometric parameters: curvature, $k_F(s)$ (full
  line), and torsion, $\tau_F(s)$ (dashed line), for the equilibrium
  solution shown in the figure \ref{fig4}.}
\label{fig5}
\end{figure}

Figure \ref{fig4} shows an example of a helix-like rod with periodic
variation of the radius of the cross section:
\begin{equation}
\label{RsHel2}
R(s)=1+0.1\sin(0.1\,s) \; .
\end{equation}
The mechanical parameters, in scaled units, are $M_3=0.04$, $\Gamma
=0.9$, $k_F(0)=0.19$, $\tau_F(0)=0.04$ and $f_3(0)=0.005$. 
The functions $k_3(s)$ and $0.1\,R(s)$ are displayed in the figure
\ref{fig4} (full line and dashed line, respectively). In this case of
periodic variation of the radius of the cross section, the
tridimensional helix-like rod displayed in figure \ref{fig4} (left) is
more complicated than that displayed in the figure \ref{fig2} (left).
Figure \ref{fig5} shows the numerical solution for the curvature
$k_F(s)$ (full line) and torsion $\tau_F(s)$ (dashed line) for the
helix-like rod shown in the figure \ref{fig4}. The curvature and
torsion are not simple functions of the arclength $s$ of the rod.

There is a kind of solution called {\it free standing helix} that is
defined by setting the resultant force
$\mathbf{F}=0$~\cite{tabor2}. In eqs. (\ref{dif}) $f_3(s)=0$ gives the
following solution for the curvature and torsion of the rod:
\begin{equation}
\label{free}
\begin{array}{ll}
(I(s)\,k_F(s))'=0 \;  \Rightarrow \; k_F(s)=\frac{k_{F_0}}{I(s)} \; , \\
\tau_F(s)=\Gamma\,k_3(s) \; \Rightarrow \; \tau_F(s)=\frac{M_3}{I(s)} 
\; .
\end{array}
\end{equation}
Notice that for the free standing helix-like rod the curvature
$k_F(s)$, and the torsion $\tau_F(s)$, will be analytical functions of
the arclength $s$ if the moment of inertia $I(s)$ is given by an
analytic function of $s$. Also $\frac{k_F(s)}{\tau_F(s)}$ is a
constant for all $s$.

The variation of the twist density along the rod can be a key factor
in a variety of phenomena. As mentioned before, the onset of a failure
in a twisted cable can be related to the increasing of the twist
density in a given region of the cable. In the Kirchhoff model, the
torsional moment $M_3$ is constant along the rod. For a relatively
high value of $M_3$ the twist density at a region of the rod with
small diameter can be so large that it may not be valid the assumption
of linear relationship between the torque and the components of the
twist vector. Also, depending on how large the moment is, the behavior
of the material could not be approximately elastic in the regions of
small diameter. So, the increase of the twist density due to the
decreasing diameter in a twisted rod may be the starting point of a
process that can culminate with its rupture.

An interesting question arises for the important phenomenon known as
{\it writhing instability}. In this phenomenon a local change in the
twist can lead to a global reconfiguration of the rod that is a
consequence of a topological constraint given by a mathematical
theorem by White~\cite{white}. If the twist density varies along the
rod, the question is to identify the region of the rod where this kind
of instability will occur. The nonhomogeneity considered here can also
have important consequences in the dynamics of this
phenomenon~\cite{golds}. It will be a subject of a future publication.

Another implication of variable twist density along twisted rods is
the problem of stability of equilibrium solutions. It is known that
above a critical value of the twist density an equilibrium solution
for the Kirchhoff model becomes unstable~\cite{goriely,tabor}. Another
good question is to investigate if a local increasing of the twist
density above the critical value, can lead to a global instability of
the related equilibrium solution. It could be important to the problem
of failure mentioned before.

A very hard problem in differential geometry is obtaining a direct
relationship between the curvature and the torsion with the radius and
the pitch of a helix-like nonhomogeneous rod. Since the definitions of
curvature and torsion involve the calculation of the modulus of the
derivatives of tangent and normal vectors with respect to the
arclength of the rod, for non constant radius and pitch, this relation
is very complicated. The analysis of this problem will be considered
in a future work.

It is interesting to note that the tridimensional configuration of the
figure \ref{fig2} displays a pattern in the radius and the pitch of
the helix-like rod seen in the spring-like tendrils of some climbing
plants. Since the young parts of the filament that composes the plant
has smaller diameter than the older parts, these filaments are
examples of rods with the nonhomogeneity considered here.

The numerical solutions obtained for $\tau_F(s)$ and the solutions for
$k_3(s)$ of the helix-like cases show that the term $\xi'$ of the
equation (\ref{hel}) is not null as it was proved to be for the case
of homogeneous helix~\cite{tyler}. It means that helix-like filaments
formed by nonhomogeneous rods are not {\it twistless}.

The existence of intrinsic curvature may lead to other planar
solutions. The helix-like solutions can also be influenced by the
intrinsic curvature. This will be considered in a more complete work.

This work was partially supported by the Brazilian agencies FAPESP,
CNPq and CAPES. The authors would like to thank Prof. Alain Goriely
for suggesting the problem.


\begin{thebibliography}{99}
\bibitem{whip} A whip is a nonhomogenous thread with varying radius of
cross section.
\end{thebibliography}
\end{document}